\documentstyle[aps,psfig,12pt]{revtex}
\newcommand{ \be }{\begin{equation}}
\newcommand{ \ee }{\end{equation}}
\newcommand{ \bea }{\begin{eqnarray}}
\newcommand{ \eea }{\end{eqnarray}}
\newcommand{ \la }{\langle}
\newcommand{ \ra }{\rangle}

\newcommand{ \eps }{{\varepsilon}}

\begin{document}

\title{ The physics of the centrality dependence of elliptic flow}

\author{S.A.~Voloshin$^{(a,b)}$ and A.M.~Poskanzer$^{(a)}$}

\address{ (a) Nuclear Science Division, Lawrence Berkeley National
  Laboratory, Berkeley, California 94720 \\
(b) Department of Physics and Astronomy, Wayne State University,
Detroit, MI 48201 }

\date{\today}

\maketitle

\begin{abstract}
The centrality dependence of elliptic flow and how it is related to
the physics of expansion of the system created in high energy nuclear
collisions is discussed.  Since in the hydro limit the centrality
dependence of elliptic flow is mostly defined by the elliptic
anisotropy of the overlapping region of the colliding nuclei, and in
the low density limit by the product of the elliptic anisotropy and
the multiplicity, we argue that the centrality dependence of elliptic
flow should be a good indicator of the degree of equilibration reached
in the reaction.  Then we analyze experimental data obtained at AGS
and SPS energies.  The observed difference in the centrality
dependence of elliptic flow could imply a transition from a hadronic
to a partonic nature of the system evolution.  Finally we exploit the
multiplicity dependence of elliptic flow to make qualitative
predictions for RHIC and LHC.

\end{abstract}


The goal of the ultrarelativistic nuclear collision program is the
creation of the QGP -- quark-gluon plasma -- the state of deconfined
quarks and gluons.  It is understood that such a state requires
(local) thermalization of the system brought about by many
rescatterings per particle during the system evolution.  It is not
clear when and if such a dynamical thermalization can really occur.
An understanding of these phenomena can be achieved by considering
elliptic flow~\cite{olli92} recently studied at AGS~\cite{e877flow2}
and SPS~\cite{na49flow} energies.  It will be shown how the centrality
dependence of the strength of elliptic flow, $v_2$, defined as the
second coefficient in the Fourier decomposition of the particle
azimuthal distribution~\cite{meth}, is an indicator of the degree of
equilibration (thermalization) achieved in the system.

Our qualitative conclusions are based on the observation, that in the
hydro limit (which we equate in our discussion to complete
thermalization) and in the opposite limiting case, the low density
limit, (where dynamical thermalization is not expected) the centrality
dependence of elliptic flow is different.  In the hydro limit, the
mean free path is much less than the geometrical size of the
system. The centrality dependence of flow is totally governed in this
case by the initial geometry (eccentricity), the latter being roughly
proportional to the impact parameter.  In the low density limit, the
mean free path is comparable to or larger than the system size.  The
final anisotropy in this case should be proportional to the ratio of
the system size to the mean free path (the number of collision).  The
anisotropy vanishes in the limit of infinite mean free path.  The
latter in its turn depends on the particle density, which is largest
for central collisions and vanishes for very peripheral collisions.
Note that the factors involved change drastically with centrality. One
could imagine other reasons for centrality dependence of elliptic flow
in the hydro model, such as the initial conditions, viscous
corrections, resonances, or effective volume corrections, but we
expect that all these other factors have a much weaker dependence on
the impact parameter. By considering the two limiting cases we hope to
highlight qualitative considerations important for understanding the
degree of thermalization and the partonic or hadronic nature of the
collisions. In essence we present a framework for examining these
questions experimentally, however at the moment, it is mainly the
experimental data which are not adequate to answer these questions
convincingly.

We will often use the term ``physics of the collision''. By this we
mean both the degree of equilibration and whether the hadronic picture
in terms of nucleons, pions, etc., or the partonic picture in terms of
deconfined quarks and gluons, is more applicable to the evolution of
the system. The partonic picture in our view is similar to a QGP but
the system is not necessarily thermalized.

{\bf Low Density Limit}

To discuss the centrality dependence of $v_2$ more quantitatively, we
start from the hypothesis that the system is {\em not} dense and its
evolution can be described by the first correction to the
collisionless limit~\cite{heisel99}.  Physically this means that the
rescattering occurring during the system evolution changes the
particle momenta very little on the average and the corresponding
change in the distribution functions can be treated in first order as
perturbations.  Under this assumption the final elliptic flow, $v_2$,
is proportional to the initial overlapping region elliptic anisotropy,
$\eps$, (introduced in flow analyses in~\cite{olli92} and in its
present form in~\cite{heisel99,sorge98}) and to the initial particle
space density which defines the probability of particles to
rescatter\cite{heisel99}.

The initial geometry of the overlapping zone can be evaluated in a
simple Glauber type model with a Woods-Saxon nuclear density.  The
results are weakly dependent on the weights used~\cite{jacobs99}.
What is important is that if one wants to compare different energies,
e.g. AGS, SPS and RHIC, the nuclear geometry cancels out, and only the
dependence on multiplicity is left.  This is true provided that the
``physics'' of of the system evolution stays the same.  If it changes
then the scaling with multiplicity will be violated.  This is a very
important point if one reads it the other way around: if scaling is
not observed then probably the physics has changed.

Under the assumption that the system is relatively dilute the momentum
anisotropy is proportional to the spatial anisotropy, but also the
particles must scatter to probe that anisotropy. Thus, the spectra
distortion is directly proportional to the spatial anisotropy and the
number of rescatterings, or the particle density in the transverse
plane.  In this limit the final elliptic flow (see a more detailed
formula in~\cite{heisel99})
\be
v_2 \propto \eps \frac{1} {S} \frac{dN}{dy},
\ee      
where $S=\pi R_x R_y$ is the area of the overlapping zone, with $R_x^2
\equiv \la x^2 \ra$ and $R_y^2 \equiv \la y^2 \ra$ describing the initial
geometrical sizes of the system in $x$ and $y$ directions,
respectively.  (The x-z axes lie in the reaction plane).  The averages
include a weighting with the number of collisions along the beam axis.
The initial space elliptic anisotropy is defined as
\be
\eps = \frac{R_y^2 - R_x^2}{R_x^2 + R_y^2}.
\ee
In our calculation we use a Woods-Saxon parameterization of the
nuclear density with parameters $R_A=1.12\cdot A^{1/3}$, and
$a=0.547$~fm.  More information on the effect of different weights and
the values of $R_x^2, R_y^2, S$ and $\eps$ as a function of impact
parameter can be found in~\cite{jacobs99}.  The proportionality
coefficient in Eq.~(1) is defined by the ``physics'' of the
rescattering.  If the physics is the same in central and peripheral
collisions then Eq.~(1) yields the centrality dependence of $v_2$.

{\bf Hydro Limit}

As follows from Eq.~(1) the elliptic flow increases with the particle
density.  Eventually it will saturate~\cite{olli99} at the hydro
limit, which would mean complete thermalization of the system.  In
this regime the centrality dependence of elliptic flow is mainly
determined by the initial elliptic anisotropy of the overlapping zone
in the transverse plane~\cite{olli99}, and the ratio of the two should
be approximately constant as shown in the first such calculations done
by Ollitrault~\cite{olli92}.  From his results it follows that
$(v_2/\eps)_{hydro}\approx 0.27-0.35$, depending on the equation of
state used (with or without QGP)\footnote{To avoid confusion, note the
difference in definitions of $\eps$ used in Eq.~(2) of this paper and
$\alpha_x$ from~\cite{olli92}.  For Pb+Pb collisions the maximal value
of $\eps \approx 0.44 $ compared to $\alpha \approx 0.3.$.  Then, the
results~\cite{olli92} yield $v_2^{\{p_t^2\}}/\eps\approx 0.55-0.7$,
where $v_2^{\{p_t^2\}}$ means the elliptic flow weighted with $p_t^2$.
Recent calculations~\cite{sollfrank} show that the particle elliptic
flow is related to this quantity as $v_2 \approx 0.5 \,
v_2^{\{p_t^2\}}$.}.  The calculations~\cite{sollfrank} give a somewhat
smaller flow, resulting in $(v_2/\eps)_{hydro}\approx 0.21-0.23$
(partly due to the realistic treatment of resonances which decrease
the pion flow by about 15\%).  Note that in both calculations,
\cite{olli92} and~\cite{sollfrank}, the longitudinal expansion of the
system is treated analytically assuming Bjorken scaling. Real 3D hydro
calculations would be very useful, although we do not expect that they
would greatly change the centrality dependence.

\vspace*{-1.0cm}
\begin{figure}
\centerline{\psfig{figure=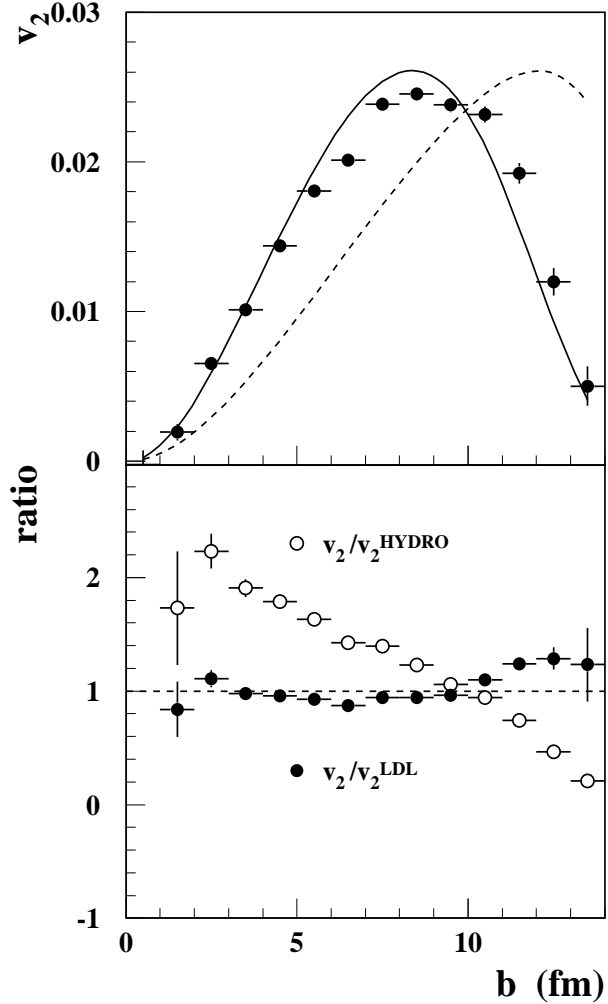,height=15.5cm}}
  \caption[]{ Top: comparison of elliptic flow, $v_2$, for pions from
  RQMD ver. 2.3 (filled circles) with the dependence expected for the
  low-density limit (solid line) and that expected for the hydro limit
  (dashed line). Bottom: ratios of $v_2/v_2^{LDL}$, and $v_2/v_2^{HYDRO}$.}
\label{fig1}
\end{figure}

{\bf RQMD}

Before discussing the experimental data we will first consider a
realistic model.  We take RQMD v2.3~\cite{sorge2.3} for our
calculations.  Fig.~1 top shows the comparison of the directly
calculated $v_2$ of pions in Pb+Pb collisions at 158~GeV$\cdot$A
collisions at mid-rapidity ($-1<y<1$) with the expectation from the
low density limit, $v_2^{LDL}$ (Eq.~(1) normalized to the same area
under the curve in order to illustrate just the centrality
dependence.)  One can see rather good agreement, which suggests that
RQMD is close to the low density limit even as one scans the
centrality from peripheral to central collisions. (In this version of
RQMD no QGP is simulated.) This is not that striking a conclusion,
considering that no hydro-type behavior has ever been observed in
RQMD. Note that the low density limit does not mean a low number of
total rescatterings. The number of rescatterings can be large provided
all of them are relatively soft and the particle momentum changes
little compared to the initial momentum. The cross section which
enters the equations is the transport (not total) cross section
(see~\cite{heisel99}).  The centrality dependence expected for the
hydro limit is shown on the same plot by a dashed line also normalized
to the same area under the curve ($v_2^{HYDRO}\approx 0.059 \eps$).
Note the large difference between the two curves, which was not noted
in~\cite{heisel99}. Fig.~1 bottom shows that the ratio of $v_2$ to the
expected functional form is flat for the low density limit but not for
the hydro limit.  A centrality dependence similar to the low density
limit was also observed in~\cite{filip} where a computer simulation of
a pion gas expansion was studied.

{\bf Data}

\begin{figure}
\centerline{\psfig{figure=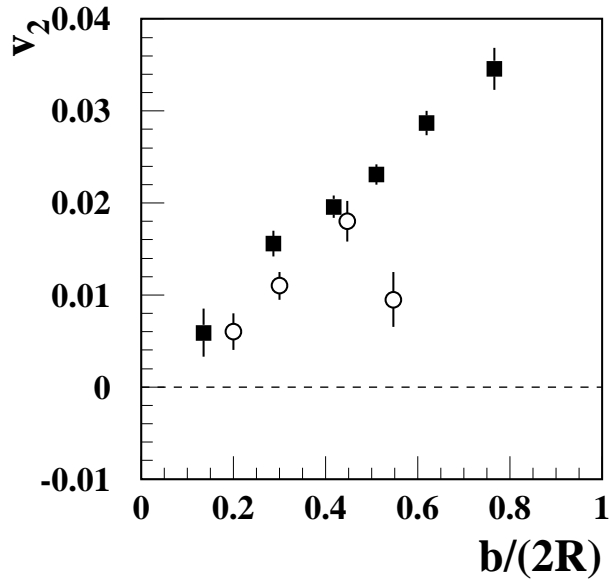,height=10.2cm}}
  \caption[]{ Elliptic flow at the AGS (open circles) and the SPS
  (filled squares). }
\label{fv2exp}
\end{figure}

Now let us turn to the experimental data.  At AGS energies the
elliptic flow of charged particles and of transverse energy was
measured by the E877 Collaboration.  Unfortunately, the
publication~\cite{e877flow2} containing the detailed pseudorapidity
dependence for each centrality lacks a figure showing just the
centrality dependence.  Our estimates based on their
data~\cite{e877flow2} of charged particle flow at midrapidity are
presented in Fig.~\ref{fv2exp}.

\vspace*{-1.2cm}

The data indicate that at AGS the flow peaks at
mid-centrality\footnote{A similar centrality dependence of transverse
energy flow (from the same data~\cite{e877flow2}) can be found in the
thesis of Chang~\cite{chang98}.}, consistent with the low density
limit prediction and no change in physics with centrality.  At this
energy some decrease of elliptic flow in peripheral collisions can be
also attributed to shadowing by spectator matter.  At
SPS~\cite{posk99}, preliminary data indicate that the elliptic flow
peak moves towards peripheral collisions.  This fact itself would hint
at the hydro-dynamical picture of the system evolution.  A more
detailed look at the data shows that this is unlikely.  First, the
maximal value of elliptic flow ($v_2 \leq 0.04$) is significantly less
than predicted by hydro calculations~\cite{olli92,sollfrank} (about
0.09--0.1)\footnote{In \cite{sollfrank} agreement was claimed between hydro
and the NA49 mid-central data~\cite{na49flow} leading to their
conclusion of complete equilibration. However, this comparison was
done for $p_t < 0.3$ GeV/c and it could be that the $p_t$ dependence
of $v_2$ in the hydro model does not agree with experiment.}. 
Second, in the
hydro limit elliptic flow should depend only on the initial space
elliptic anisotropy, $\eps$.  The preliminary NA49 data indicate that
the ratio $v_2/\eps$, at least for semi-central collisions, is likely
increasing with centrality~\cite{posk99} (see the data presented in
Fig.~3 below).  
This centrality dependence
(natural for the low density limit) implies that we still could be far
from the hydro regime\footnote{
One can argue that, taking into
account systematic uncertainties, the preliminary SPS data for
$v_2/\eps$ are consistent with being constant as a function of
centrality. In this case it would indeed mean that the system has
equilibrated and the hydro regime has been reached.  The low absolute
strength of the elliptic flow in this case would indicate that the
equilibration happens at a rather late time when the spatial
anisotropy $\eps$ has decreased due to initial ``free streaming''.  We
do not exclude this possibility but must wait for the final SPS data
and the coming RHIC data to answer the question.}. 

Assuming that at SPS the hydro regime is not reached yet, the observed
centrality dependence of elliptic flow would indicate that the
physics of the system evolution is different in central and peripheral
collisions.  
Elliptic flow peaks at more peripheral collisions because 
the central collisions  exhibit too little flow compared to that
expected from the AGS data scaled with multiplicity. 
A natural explanation for this would be that peripheral
collisions are described by hadronic (re)scatterings (the same as at
the AGS in both peripheral and central collisions) while in central
collisions partonic physics becomes important.
One of the possible mechanisms responsible for the change could be a
color percolation occurring at high parton densities in the central
collisions and discussed in more detail below.

{\bf Discussion}

Summarizing, our view of the overall picture is: at AGS energies, the
physics of rescattering which defines the system evolution is hadronic
in nature, while at SPS it is the same for peripheral collisions, but
for central collisions the physics is likely to be partonic.  The
partonic picture will remain at RHIC energies, with some extension
toward more peripheral collisions.  At RHIC equilibration becomes more
important, but it is not clear if complete thermalization will be
reached.  At LHC energies the parton densities could become so high
that (partonic) rescattering would lead to dynamical equilibration of
the (partonic) system (creation of regions of real QGP) and
consequently to a hydro-dynamical type of system evolution.

The above picture for collisions of heavy nuclei implies that the
shape of the centrality dependence of elliptic flow would change
continuously with beam energy. 
At AGS, the elliptic flow is peaked at an impact parameter value
slightly higher than $R_A$, just as prescribed by the low density
limit.
At SPS energies the peak moves toward more peripheral collisions because
possibly the physics of relatively central collisions may have changed from
hadronic to partonic, which leads to weaker flow than one would expect
taking into account the increased multiplicity.  If thermalization is
not reached at RHIC, the elliptic flow peak could move back toward mid-central
collisions because the physics of the peripheral and central
collisions will be the same -- partonic rescattering, unlike the
situation at SPS when peripheral collisions are driven by the hadronic
rescatterings resulting in relatively large flow signal.  
At even higher
energies at LHC, the elliptic flow should peak at more peripheral
collisions just as predicted by hydrodynamic calculations.

\begin{figure}
\centerline{\psfig{figure=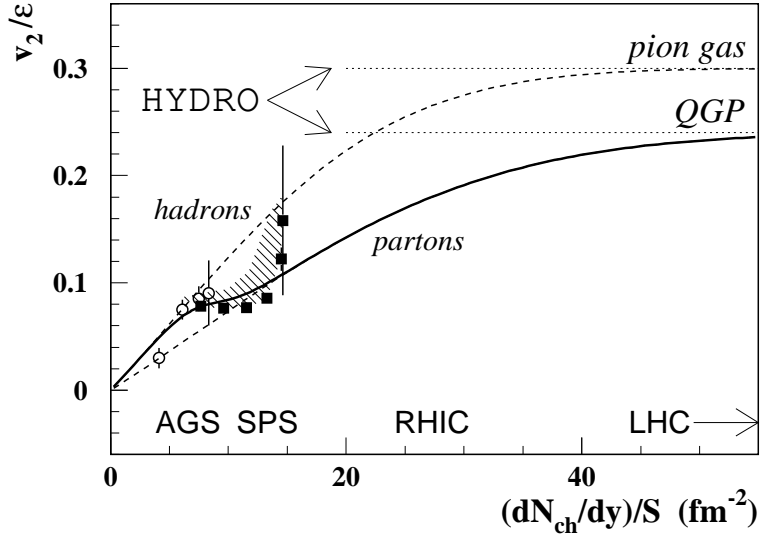,height=9.0cm}}
  \caption[]{ Elliptic flow divided by the initial space elliptic
  anisotropy at the AGS (open circles) and the SPS (filled
  squares). The shaded area shows the uncertainty in the SPS
  experimental data due to the uncertainty in the centrality
  determination.  See text and footnote for the description of the
  curves and hydro limits.}
\label{fall}
\end{figure}

The schematic overall picture based on these observations is presented
in Fig.~\ref{fall}, where the ratio of elliptic flow to the initial
space elliptic anisotropy is presented as a function of initial
particle density.\footnote{At the moment this plot is qualitative as
many things shown have large uncertainties.  The hydro limits can
depend slightly on the initial particle
density~\cite{olli92,sollfrank} and, more importantly, on the time of
thermalization of the system.  The values shown are an average of the
results of~\cite{olli92,sollfrank}.  The predictions for the case
without QGP are only for the EoS of a massless pion gas. Resonances
can soften the EoS and lead to weaker flow.  The uncertainty in the
experimental points is mainly from the determination of the collision
centrality required for calculation of the initial space elliptic
anisotropy and the area of the overlapping region.  The data points
correspond to the centrality determined from the fraction of the total
cross section corresponding to each centrality bin. Higher
centralities were estimated from experimental measurement of the
number of participants~\cite{cooper99}. Finally, the smooth dashed
curves are just schematic illustrations for hadronic and partonic
scenarios and the solid curve includes a transition between the two.}
In this plot we use the experimental charged particle multiplicity,
assuming that it is proportional to the total particle multiplicity
and also to the initial particle multiplicity.  For the experimental
values we use $dN_{ch}/dy$ at mid-rapidity
from~\cite{cooper99,e877dndeta}.
            
\vspace*{-1.0cm}

In the limit of very low density the objects which rescatter must be
hadrons.  At some critical density a partial deconfinement happens.
Parton density becomes high enough such that the color parton can
propagate in the perpendicular plane without hadronization.  
Each parton is always close
enough to other partons which screen its color\footnote{This picture
is very close to the deconfinement (color percolation) model discussed
by Satz~\cite{satz99} for $J/\Psi$ suppression.}.  Once the motion in
the perpendicular plane becomes easier (there is no need 
for hadronization), the elliptic flow decreases.
Note that the system still can be far from being dynamically
thermalized, which would occur only at even higher particle densities.
Even more important, such a significant change in the behavior of
$v_2/\eps$ can only happen if the system is not thermalized.  See also
the discussion of this question in~\cite{sorge98,sollfrank} along with
the discussion of the possibility of observation of the QGP to hadron
gas phase transition.

To prove or disapprove the picture described above one needs more
accurate data on the centrality dependence of elliptic flow.  We would
like to emphasize the importance of flow measurements not only at
medium impact parameters but in the full range of centrality including
rather central collisions where the anisotropic flow is small.  The
measurement of elliptic flow and its centrality dependence at RHIC
thus becomes very important.  Different models predict different
rapidity densities for RHIC and LHC. Assuming that they are higher
than at SPS by factors of 2 and 8, respectively, we have indicated the
regions expected for Au+Au (Pb+Pb) collisions in Fig.~\ref{fall}.  The
measurements of elliptic flow in collisions of lighter systems
(e.g. Cu+Cu) are also very important since they would cover the region
of the SPS Pb+Pb data and would be useful in testing the above
picture.  The new SPS data taken at 40~GeV$\cdot$A energy are also of
great interest since they would bridge the two other sets of data and
may scan the onset of deconfinement from hadronic to partonic physics.

Note that our picture of nuclear collisions and QGP production is
different from what is usually discussed, which assumes thermal
equilibrium even at rather low beam energies, when QGP is not
expected, and then with an increase in collision energy, formation of
regions of QGP.  We believe that what could happen is that the
deconfinement can occur before dynamical thermalization is
achieved~\cite{stock99} and that the centrality dependence of elliptic
flow would be a good indicator of this.  

We are grateful to J.-Y.~Ollitrault, U.~Heinz, H.~Heiselberg,
G.~Cooper, P.~Seyboth, R.~Snellings, H.~Sorge, and H.G.~Ritter for
useful discussions.

This work was supported by the Director, Office of Energy Research,
Office of High Energy and Nuclear Physics, Division of Nuclear Physics
of the U.S. Department of Energy under Contracts DE-AC03-76SF00098
and DE-FG02-92ER40713.
  


\end{document}